\documentclass[twocolumn]{aastex6}
\usepackage{float}
\usepackage{hyperref}

\slugcomment{To appear in JAA special issue   on RECTO-III.}
\shorttitle{High Energy Transients: the Millisecond Domain }
\shortauthors{Rao}

\begin{document}

\title{High Energy Transients:  the Millisecond Domain }

\altaffiltext{1}{Email: \url{arrao@tifr.res.in}}
\altaffiltext{2}{Tata Institute of Fundamental Research, Mumbai, India}
%\altaffiltext{3}{Inter University Centre for Astronomy \& Astrophysics, Pune, India}
%\altaffiltext{4}{Nicolaus Copernicus Astronomical Center, Polish Academy of Sciences, Warsaw, Poland}
%\altaffiltext{5}{Vikram Sarabhai Space Centre, Thiruvananthapuram, India}
%\altaffiltext{6}{Physical Research Laboratory, Ahmedabad, India}
%\altaffiltext{7}{ISRO Satellite Centre, Bengaluru, India}

\author{A. R. Rao\altaffilmark{1,2}} %\fnref{rao}
%\author{M. H. Patil\altaffilmark{2}}
%\author{Yash Bhargava\altaffilmark{2,3}}
%\author{Rakesh Khanna\altaffilmark{2}}
%\ \author{D. Bhattacharya\altaffilmark{3}}
%\author{V. Bhalerao\altaffilmark{3}}
%\author{S. V. Vadawale\altaffilmark{6}}

\begin{abstract}
Search for high energy transients in the millisecond domain has come to the focus in recent times due to the detection of Gravitational Wave events and the identification of Fast Radio Bursts as cosmological sources. I will highlight the  sensitivity limitations in the currently operating hard X-ray telescopes and give some details of the search for millisecond events in the AstroSat CZT Imager data.

\end{abstract}


\section{Introduction}
\label{intro}

Gamma-ray bursts (GRBs) are the archetypical transients in the hard X-ray domain: when they occur
they are the brightest objects in the sky, they are so bright that even a simple small area 
detector can detect them, they are intrinsically so luminous that we can see GRBs into far corners
of the universe   \citep[see][for a review of GRBs]{Kumar2015}.
Observationally, GRBs span about four orders of magnitude in flux (10$^{-7}$ - 10$^{-3}$ erg cm$^{-2}$) and
similar order of magnitude in time ($\sim$10 ms to $>$ 100 s). At the fainter flux level, the GRB flux distribution
is very flat \citep{Fishman1995} - there are not many faint GRBs. GRBs are traditionally separated into short and long GRBs
and the short GRBs peak at around 1 s \citep{Berger2014} and the number of GRBs with duration  much less than a second is
quite low. Further, GRBs have a low occurrence rate: the estimated all sky rate is about one per day \citep{Kumar2015}.

  The question, therefore, is whether there are population of transients in the high energy domain
  with properties drastically different from those of GRBs: like, say, in the domain spanned by
  lower flux level, shorter time scales and higher occurrence rates. This is particularly relevant due
  to some new inputs in recent times: the detection of Gravitational Wave (GW) events and the 
  discovery of Fast Radio Bursts (FRBs).
  
The first GW event  discovered on 2015    September 14 \citep{Abbott2016} opened up the new and fascinating
field of merging black holes. Finding electromagnetic counterparts of GW events is one of the major
research goals: it is quite challenging due to the large positional inaccuracy of GW events.
Hence, the report by \citet{Connaughton2016} that GW 150914 was accompanied, although 0.4 s later than the event,
by a hard X-ray burst detected by the Gamma-ray Burst Monitor (GBM) onboard the Fermi satellite
was seen as a breakthrough result. The authors claimed a 5.1 sigma detection with a 
false alarm probability of less than 1\%. The same data, however, was reanalysed by \citet{Greiner2016} and
they concluded that this burst was most likely a background fluctuation rather than an astrophysical 
event. This result highlights the difficulty of finding faint bursts by GRB monitors: by design they are
open all-sky monitors and hence prone to background fluctuations and false events induced by
the omnipresent Cosmic Rays. 

The discovery of Fast Radio Bursts  \citep[see][for a review]{Katz2016}  demonstrates  that real surprises are 
round the corner when   unexplored parameter regions are observationally explored. Unlike GRBs, FRBs are
much more common (estimated rates are several thousand per day), they last for very short durations (milliseconds) and,
so far, are confined only to the radio band of the electromagnetic radiation. These new discoveries 
of recent times (GW events and FRBs) compels one to explore high energy region in new unexplored regimes:
shorter duration and fainter fluxes. It is argued here that CZT Imager (CZTI) of the AstroSat satellite
has several new and fascinating features which will make it an ideal instrument to look for
fainter hard X-ray events in the millisecond domain. In the next section an overview of the difficulties
of hard X-ray observations are highlighted and in \S 3, the design characteristics of CZT Imager which makes
it a sensitive hard X-ray monitor is described. We also give the recent results of CZTI as a GRB monitor 
and highlight its efficacy as a sensitive instrument to search for hard X-ray bursts associated with
GW events and FRBs.

\section{The hard X-ray domain}
\label{hxray}

The hard X-ray band, the region where the emission is dominated by non-thermal processes like
synchrotron radiation and  inverse Compton scattering,  is the band which probes Astrophysical sites
exhibiting  exotic phenomena like accretion onto black holes, jet launching and the mysterious GRBs possibly
signalling the birth of black holes.  Observationally, the X-ray focussing techniques have achieved very high
sensitivity for narrow field of view observatories like the NuSTAR satellite \citep{Harrison2013} upto about 80 keV. But, for all sky monitoring in this
region, particularly above 80 keV, the available sensitivity of detectors is quite modest. For example,
in this energy range the first all sky survey was conducted by the HEAO-A satellite, launched in 1977. In two
years of operation only  22 sources were recorded \citep{Levine1984}. Two of the most successful hard X-ray 
detectors of recent times fared better: the Swift/BAT detector (launched in 2004) recorded 86 sources
above 80 keV in the first 6 years of its operation \citep{Cusumano2010} and the Intergral/IBIS instrument, launched in
2002, recorded 132 sources in its first 11 years of operation \citep{IBIS2010}. 

It can be noticed that in spite of all the  advances of recent times in sophistications in detector technology and 
improvements in space hardware fabrication, the increased number of sources detected above 
80 keV is rather a reflection of the increased duration of observation than any improvements
in sensitivity. The major reason behind this is the large and fluctuating background
in space environment \citep[see][for a discussion on space background in hard X-ray
and gamma-ray regions]{DeanBack1991} for a discussion on space background in hard X-ray
and gamma-ray regions). 

The CZT-Imager (CZTI) of AstroSat has several innovative and new design features specifically
implemented to improve the sensitivity above 80 keV. In the next section we discuss these special
design features and describe the utility of CZTI as a sensitive hard X-ray monitor.

\section{CZT-Imager onboard AstroSat}
\label{czti}

The $AstroSat$ satellite is a  multi-wavelength astronomical observatory and it  was launched on 2015 September 28.
It includes three co-aligned X-ray instruments:  the Soft X-ray Telescope (SXT),  Large Area X-ray Proportional Counters (LAXPCs), and the Cadmium-Zinc-Telluride Imager (CZTI). Additionally,  the Ultra-Violet Imaging Telescope (UVIT) provides a deep and wide field image of the
sky and the Scanning Sky Monitor (SSM) observes about  half the celestial  sphere for X-ray transients  \citep{2014SPIE.ASTROSAT}. 
The orbit of $AstroSat$ is selected specifically for very  low background for X-ray detectors:
it has an 
  altitude of 650 km in a nearly equatorial circular orbit (inclination of $ 6^{\circ} $).

  The primary design considerations of CZT Imager was to have an area and sensitivity
   comparable to
  the recent/ current  best hard X-ray telescopes like HEXTE onboard RXTE, BAT onboard 
  Swift or IBIS onboard INTEGRAL. The hard X-ray observations were desiged to complement
  the SXT and LAXPC data to provide continuum X-ray spectroscopy in an extremely wide band width of
  0.3 to 150 keV. Additionally, CZT Imager was built with a weight of about 50 kg and typical size
  of about 60 cm. This needs to be compared with weight and size of INTEGRAL 
  (2000 kg and 500 cm) and Swift (1500 kg, 560 cm), respectively. 
  
  Hard X-ray detectors are generally very heavy because of the need of the use of
  heavy elements to block the off-axis X-ray and gamma-rays. This shield, sometimes generates
  its own characteristic X-rays and to suppress these additional material of different atomic number
  is used in a what is called the graded-shield configuration \citep{DeanShield1976}.  In the space environment, however,
  the omnipresent Cosmic Rays induce background in these very shields to increase the background \citep{DeanBack1991}.
  To alleviate these problems, some special design features are introduced in CZTI. To start with,
  the shield is designed only for low energies (less than 100 keV) such that the weight is drastically
  reduced. The trade-off is the slightly inferior sensitivity above 100 keV for sources being targeted. This,
  however, is more than compensated by the fact that CZTI acts as a true all-sky monitor above these
  energies.

    The low Inclination (6$^\circ$) of the $AstroSat$ orbit offers stable and low 
    background. In addition to this, CZTI uses pixelated semi-conductor devices
    arranged in a modular fashion and the facility to transfer individual photon data (correct to 20 micro-sec)
    is
    extremely useful for using sophisticated off-line software for noise reduction.
    Further, a coded aperture mask enables the simultaneous measurement of
    background. In Figure 1, the configuration of CZTI is shown and more technical details
    can be found in \citet{BhaleraoCZTI}.

\begin{figure*}[!ht]
\centering
\includegraphics[width=0.5\paperwidth]{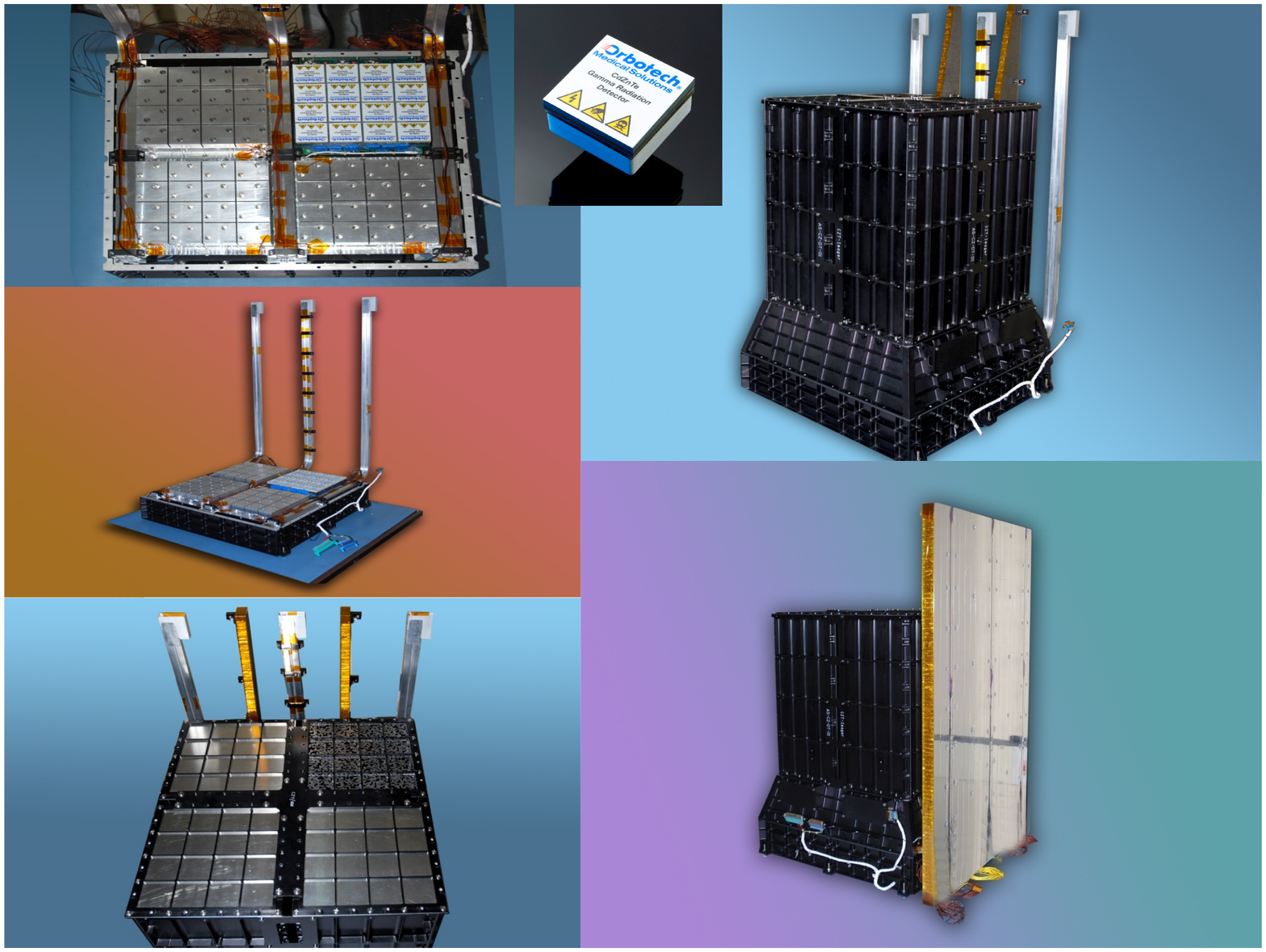}
\caption{CZT Imager is built from a mosaic of 64 Orbotech detector modules (shown as an inset at
the top), arranged in 4 identical quadrants (top left). An elaborate cooling arrangement with heat pipes is used
to keep the detectors at a cntrolled tremperature of 5 - 10$^\circ$ C (left middle and bpottom). The
assembled CZT Imager (right middle and bottom) with an external radiator plate is built with a 
total weight of about 50 kg.}
\label{fig:czti}
\end{figure*}\par

\subsection{CZT-Imager as a GRB monitor}
\label{czti-grb}

AstroSat CZT Imager detected  GRB 151006A on the first day of operation. 
This GRB was 
incident at 60$^\circ$.7 from vertical
       ($\theta_x$  = 34$^\circ$; $\theta_x$  = 58$^\circ$)
 \citep{Rao2016}. 
GRB 151006A was extensively studied combining data from Fermi-GBM and Swift-BAT 
\citep[see also][]{Basak2017}. This is a 
peculiar GRB with peak energy  of 2 MeV.
Joint spectral fitting with GBM, BAT, CZTI and CZTI-Veto
demonstrated that CZTI with Swift-BAT can provide spectral results comparable to
that obtained from Fermi.
It was also demonstrated that CZTI can also provide coarse localisation.
CZTI along with Fermi and Swift, currently provide complementary information on
GRBs. In the  first year  of operation, there were a total of    214 GRBs detected 
by various satellite missions (about 
 150  from Fermi and about  60 from Swift). From a targeted search  about 40 - 50 
 GRBs are found in the CZTI data during this period. A rigorous software to look for GRBs in the CZTI data
 is being developed and it is envisaged that CZTI should detect about 100 - 150 GRBs per year.
 A full AstroSat mass model  is generated to enable a formal and proper localisation
 of CZTI detected GRBs.

CZTI has best sensitivity to GRBs in 150 to 400 keV energy range and one of the
most important and exciting capability of CZTI is its ability to measure the
hard X-ray (150 - 400 keV) polarisation of bright GRBs \citep{Tanmoy2017}. 
A systematic analysis of the 11 brightest  GRBs detected by CZTI during its  first year of operation yielded significant
polarisation measurements. Significant hard X-ray  polarisation was measured in  seven of the 11 GRBs and 
meaningful upper limits could be placed for the remaining four GRBs. This number effectively doubles the number of GRBs with 
measured hard X-ray polarisation \citep{Toma2009}.

\begin{figure*}[!ht]
\centering
\includegraphics[width=0.3\paperwidth]{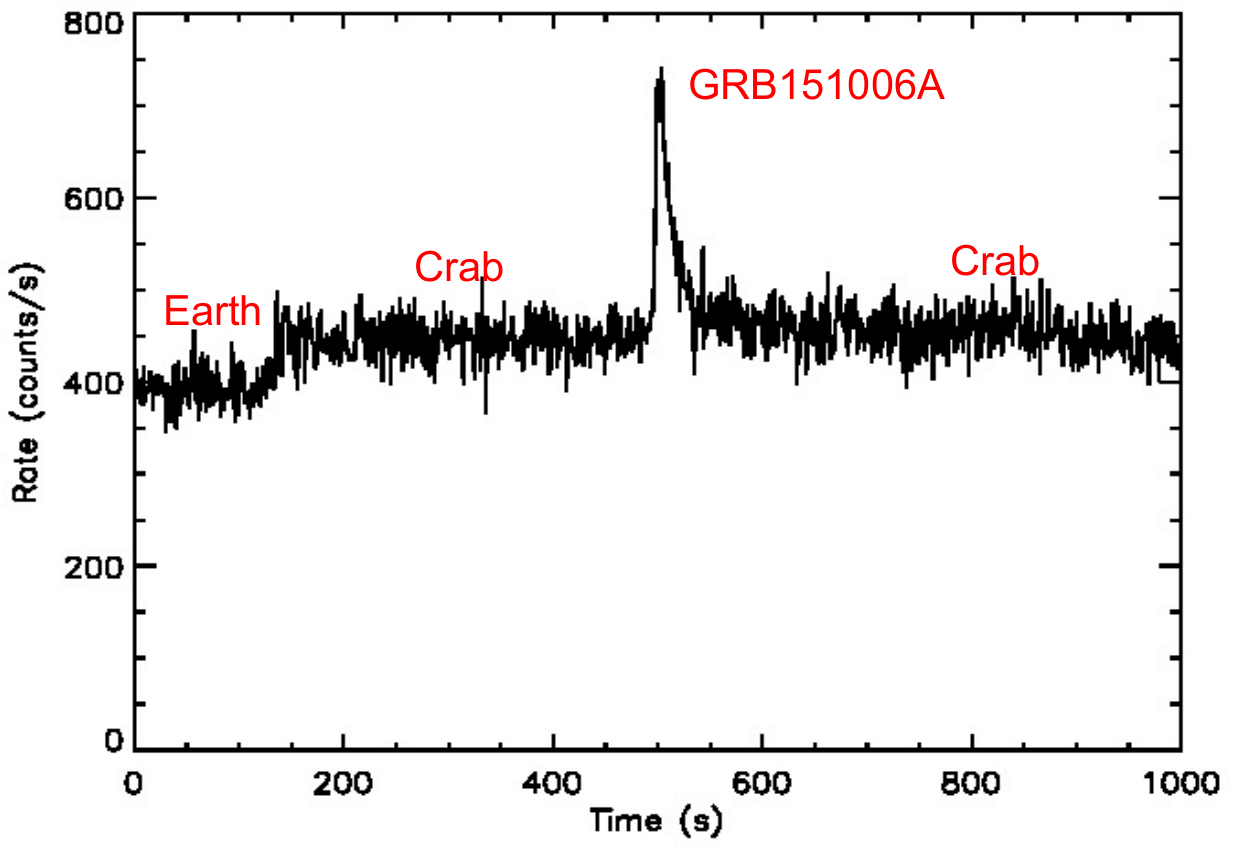}
\includegraphics[width=0.3\paperwidth]{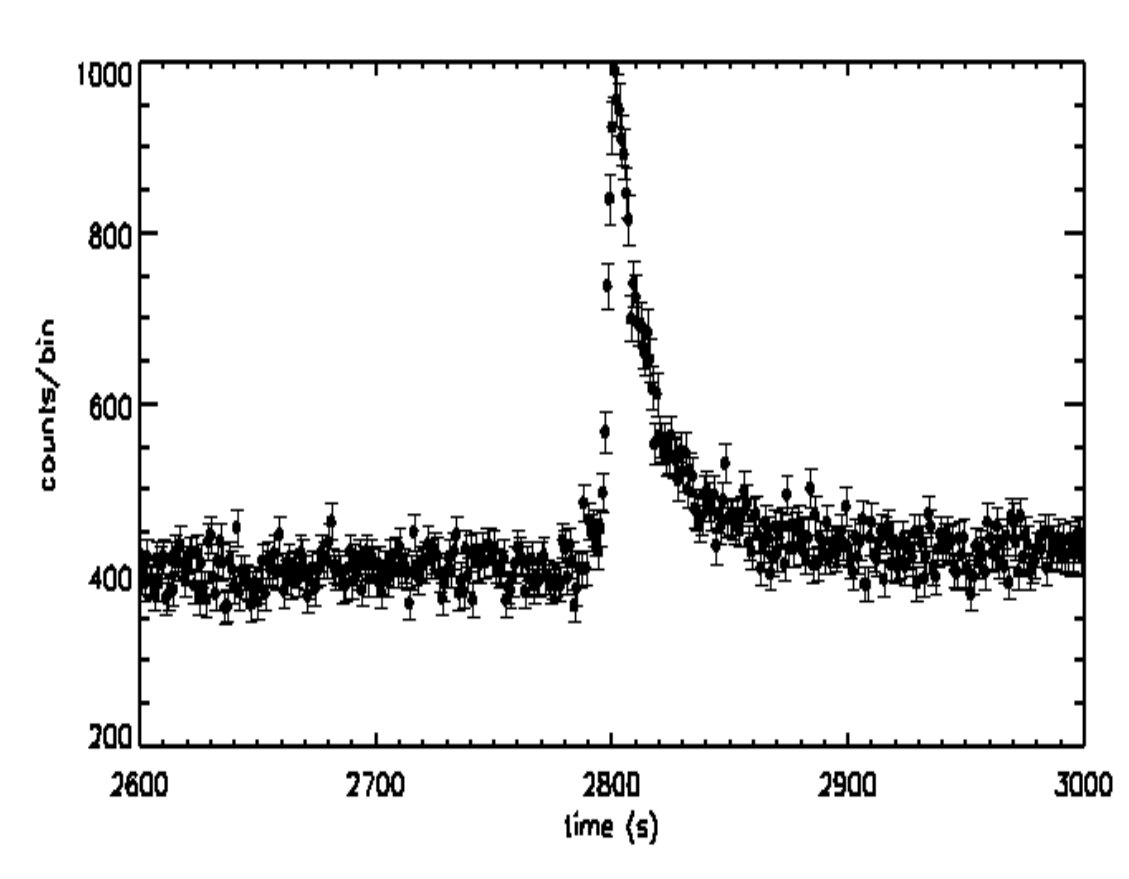}
\includegraphics[width=0.3\paperwidth]{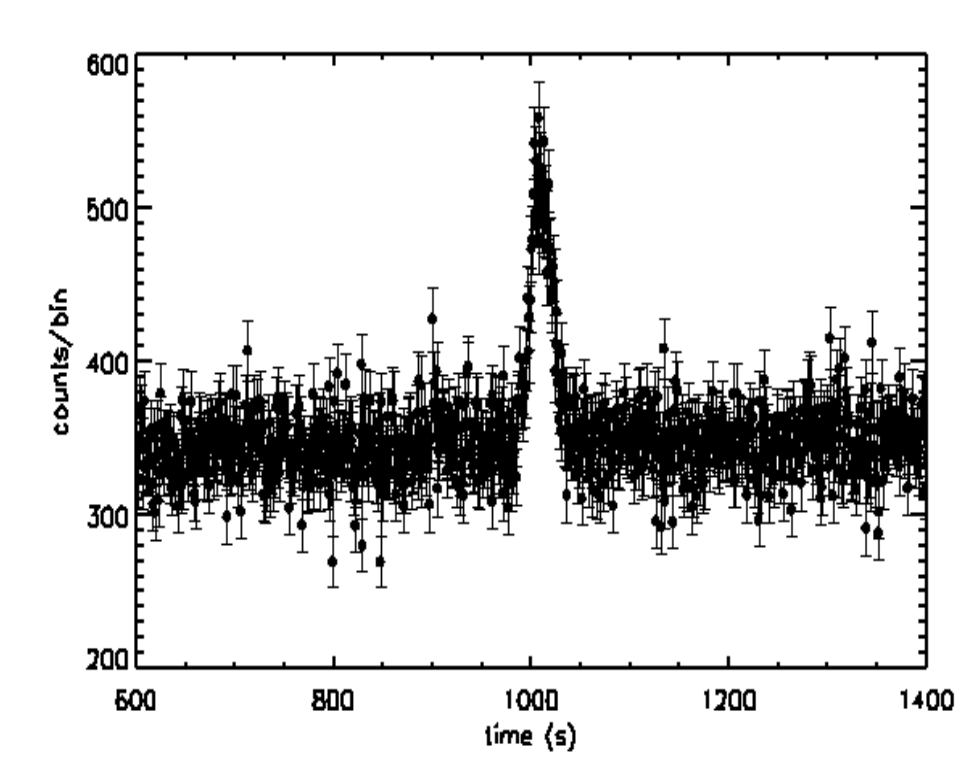}
\includegraphics[width=0.3\paperwidth]{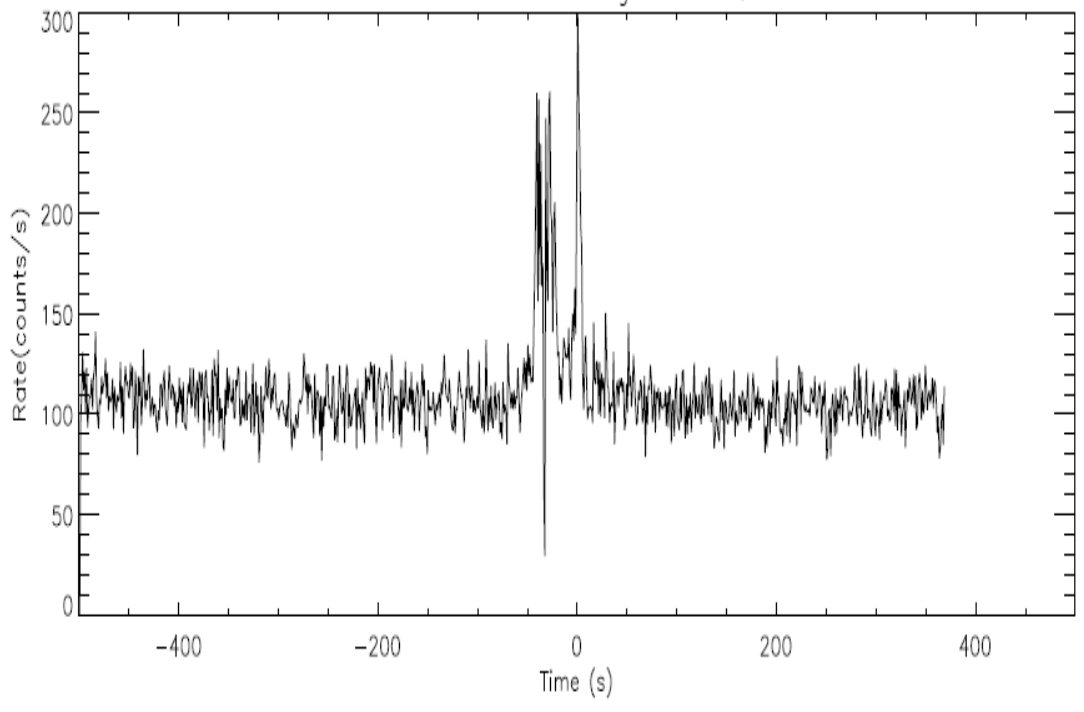}
\caption{Light curves of a few representative GRBs as seen by CZT Imager. During the
first day of operation GRB 151006A was seen along with Crab observation (top left). GRB 160131A (top right) has a FRED type profile whereas 
  GRB  160119A (bottom left) shows a single symmetric peak. GRB 160325A (bottom right) happens 
  to shine through the 4.$^\circ$8 field of view of the detector and shows saturation
  peaks in the light curve. }
\label{fig:grb}
\end{figure*}\par

\section{CZT-Imager and GW events }
\label{czti-gw}

The detection of GW events on 2015 September 14 \citep{Abbott2016} opened up the new and exciting
field of GW astrophysics. Measuring the electromagnetic counterparts of GW events is
one of the crucial and path breaking research in the present day high energy astrophysics.
This first GW event was thought to be associated with a gamma-ray counterpart \citep{Connaughton2016}, which, however,
could not be corroborated by further rigorous statistical tests for association \citep{Greiner2016}. This
emphasises the need for localisation (even at the level of a few degrees) for faint 
transients. 
 With its wide field of view and higher sensitivity and some limited localisation
 capability, CZT Imager can provide very exciting inputs in this field.
 
 The association of the GW event  GW170104 and GRB170105A is a case in point
 \citep{Bhalerao2T}.   GW event  was reported by the LIGO-Virgo collaboration (triggered 
 on 2017 Jan 4 at 10:11:58.599 UTC) and the  localisation accuracy 
 spanned about a thousand square degrees \citep{Bhalerao2T}. Looking for associated transients
 in such large error regions in the sky, however, was a challenge. The ATLAS collaboration
 reported a  object called  ATLAS17aeu: a fading optical object from the same 
 general direction as GW170104.  This object was thoroughly studied and
 the GROWTH collaboration made a fit to the exponentially falling
 light curve and derived a reasonably robust start time of this optical event and
 found that the start time is delayed as compared to the GW event by as much as
 21.5$\pm$1.0 hours \citep{Bhalerao2T}.  The $AstroSat$ CZTI data was analysed thoroughly and it
 was found that at the time of this extrapolated start time there indeed
 was a gamma-ray burst, named GRB 170105A. \citet{Bhalerao2T} give a detailed
 analysis of these results and conclude that  GRB 170105A is
 unrelated but   fortuitously having a spatial coincidence with GW170104. 
 The CZTI data could provide a rough localisation which helped in the
 association of the afterglow with the GRB. CZTI could also provide
 flux upper limits during the GW event, which is the most stringent in the 0.1 s time scale.

For short events CZTI has excellent sensitivity  (about 10$^{-7}$ erg cm$^{-2}$  s$^{-1}$) and
it can be improved upon by a more careful data screening. In the unexplored region of
millisecond regime, CZTI can provide good data which will be useful for
GW searches, short GRBs and exploring new regions of parameter space.

\section{Conclusions }
\label{concl}

CZT Imager of the $AstroSat$ satellite  is proving to be a good monitor for GRBs above 100 keV. It 
has the best sensitivity in the 100 Ð 300 keV region (most of the GRBs have peak energy here). By using the
satellite structure as a coder, it has 
some localisation capability  for bright GRBs. The availability of 
individual photon counting and position information is very crucial for understanding the
systematics in the data and it will be very useful to identify  very faint short events.
A systematic analysis for all events is going on and it is expected to provide very
critical information on GRBs like polarisation, search for faint short events etc.

\par

\section*{Acknowledements}

This publication uses the data from the AstroSat mission of the Indian Space Research Organisation (ISRO), archived at the Indian Space Science Data Centre (ISSDC). CZT-Imager is built by a consortium of Institutes across India including Tata Institute of Fundamental Research, Mumbai,  Vikram Sarabhai Space Centre, Thiruvananthapuram, ISRO  Satellite Centre, Bengaluru,  Inter University Centre for Astronomy and Astrophysics, Pune, Physical Research Laboratory, Ahmedabad, Space Application Centre, Ahmedabad: contributions from the vast technical team from all these institutes are gratefully acknowledged.

\bibliographystyle{apj}

\bibliography{ref}

\begin{thebibliography}{}
\expandafter\ifx\csname natexlab\endcsname\relax\def\natexlab#1{#1}\fi

\bibitem[{{Abbott} {et~al.}(2016){Abbott}, {Abbott}, {Abbott}, {Abernathy},
  {Acernese}, {Ackley}, {Adams}, {Adams}, {Addesso}, {Adhikari}, \&
  et~al.}]{Abbott2016}
{Abbott}, B.~P., {Abbott}, R., {Abbott}, T.~D., {et~al.} 2016, Physical Review
  Letters, 116, 241103

\bibitem[{{Basak} {et~al.}(2017){Basak}, {Iyyani}, {Chand}, {Chattopadhyay},
  {Bhattacharya}, {Rao}, \& {Vadawale}}]{Basak2017}
{Basak}, R., {Iyyani}, S., {Chand}, V., {et~al.} 2017, ArXiv e-prints,
  arXiv:1707.09924

\bibitem[{{Berger}(2014)}]{Berger2014}
{Berger}, E. 2014, \araa, 52, 43

\bibitem[{{Bhalerao} {et~al.}(2017{\natexlab{a}}){Bhalerao}, {Kasliwal},
  {Bhattacharya}, {Corsi}, {Aarthy}, {Adams}, {Blagorodnova}, {Cantwell},
  {Cenko}, {Fender}, {Frail}, {Itoh}, {Jencson}, {Kawai}, {Kong}, {Kupfer},
  {Kutyrev}, {Mao}, {Mate}, {Mithun}, {Mooley}, {Perley}, {Perrott}, {Quimby},
  {Rao}, {Singer}, {Sharma}, {Titterington}, {Troja}, {Vadawale}, {Vibhute},
  {Vedantham}, \& {Veilleux}}]{Bhalerao2T}
{Bhalerao}, V., {Kasliwal}, M.~M., {Bhattacharya}, D., {et~al.}
  2017{\natexlab{a}}, \apj, 845, 152

\bibitem[{{Bhalerao} {et~al.}(2017{\natexlab{b}}){Bhalerao}, {Bhattacharya},
  {Vibhute}, {Pawar}, {Rao}, {Hingar}, {Khanna}, {Kutty}, {Malkar}, {Patil},
  {Arora}, {Sinha}, {Priya}, {Samuel}, {Sreekumar}, {Vinod}, {Mithun},
  {Vadawale}, {Vagshette}, {Navalgund}, {Sarma}, {Pandiyan}, {Seetha}, \&
  {Subbarao}}]{BhaleraoCZTI}
{Bhalerao}, V., {Bhattacharya}, D., {Vibhute}, A., {et~al.} 2017{\natexlab{b}},
  Journal of Astrophysics and Astronomy, 38, 31

\bibitem[{{Chattopadhyay} {et~al.}(2017){Chattopadhyay}, {Vadawale}, {Aarthy},
  {Mithun}, {Chand}, {Basak}, {Rao}, {Mate}, {Sharma}, {Bhalerao}, \&
  {Bhattacharya}}]{Tanmoy2017}
{Chattopadhyay}, T., {Vadawale}, S.~V., {Aarthy}, E., {et~al.} 2017, ArXiv
  e-prints, arXiv:1707.06595

\bibitem[{{Connaughton} {et~al.}(2016){Connaughton}, {Burns}, {Goldstein},
  {Blackburn}, {Briggs}, {Zhang}, {Camp}, {Christensen}, {Hui}, {Jenke},
  {Littenberg}, {McEnery}, {Racusin}, {Shawhan}, {Singer}, {Veitch},
  {Wilson-Hodge}, {Bhat}, {Bissaldi}, {Cleveland}, {Fitzpatrick}, {Giles},
  {Gibby}, {von Kienlin}, {Kippen}, {McBreen}, {Mailyan}, {Meegan}, {Paciesas},
  {Preece}, {Roberts}, {Sparke}, {Stanbro}, {Toelge}, \&
  {Veres}}]{Connaughton2016}
{Connaughton}, V., {Burns}, E., {Goldstein}, A., {et~al.} 2016, \apjl, 826, L6

\bibitem[{{Cusumano} {et~al.}(2010){Cusumano}, {La Parola}, {Segreto},
  {Ferrigno}, {Maselli}, {Sbarufatti}, {Romano}, {Chincarini}, {Giommi},
  {Masetti}, {Moretti}, {Parisi}, \& {Tagliaferri}}]{Cusumano2010}
{Cusumano}, G., {La Parola}, V., {Segreto}, A., {et~al.} 2010, \aap, 524, A64

\bibitem[{{Dean} {et~al.}(1991){Dean}, {Lei}, \& {Knight}}]{DeanBack1991}
{Dean}, A.~J., {Lei}, F., \& {Knight}, P.~J. 1991, \ssr, 57, 109

\bibitem[{{Dean} \& {Nikiforidis}(1976)}]{DeanShield1976}
{Dean}, A.~J., \& {Nikiforidis}, G. 1976, \aap, 52, 409

\bibitem[{{Fishman} \& {Meegan}(1995)}]{Fishman1995}
{Fishman}, G.~J., \& {Meegan}, C.~A. 1995, \araa, 33, 415

\bibitem[{{Greiner} {et~al.}(2016){Greiner}, {Burgess}, {Savchenko}, \&
  {Yu}}]{Greiner2016}
{Greiner}, J., {Burgess}, J.~M., {Savchenko}, V., \& {Yu}, H.-F. 2016, \apjl,
  827, L38

\bibitem[{{Harrison} {et~al.}(2013){Harrison}, {Craig}, {Christensen},
  {Hailey}, {Zhang}, {Boggs}, {Stern}, {Cook}, {Forster}, {Giommi},
  {Grefenstette}, {Kim}, {Kitaguchi}, {Koglin}, {Madsen}, {Mao}, {Miyasaka},
  {Mori}, {Perri}, {Pivovaroff}, {Puccetti}, {Rana}, {Westergaard}, {Willis},
  {Zoglauer}, {An}, {Bachetti}, {Barri{\`e}re}, {Bellm}, {Bhalerao},
  {Brejnholt}, {Fuerst}, {Liebe}, {Markwardt}, {Nynka}, {Vogel}, {Walton},
  {Wik}, {Alexander}, {Cominsky}, {Hornschemeier}, {Hornstrup}, {Kaspi},
  {Madejski}, {Matt}, {Molendi}, {Smith}, {Tomsick}, {Ajello}, {Ballantyne},
  {Balokovi{\'c}}, {Barret}, {Bauer}, {Blandford}, {Brandt}, {Brenneman},
  {Chiang}, {Chakrabarty}, {Chenevez}, {Comastri}, {Dufour}, {Elvis}, {Fabian},
  {Farrah}, {Fryer}, {Gotthelf}, {Grindlay}, {Helfand}, {Krivonos}, {Meier},
  {Miller}, {Natalucci}, {Ogle}, {Ofek}, {Ptak}, {Reynolds}, {Rigby},
  {Tagliaferri}, {Thorsett}, {Treister}, \& {Urry}}]{Harrison2013}
{Harrison}, F.~A., {Craig}, W.~W., {Christensen}, F.~E., {et~al.} 2013, \apj,
  770, 103

\bibitem[{{Katz}(2016)}]{Katz2016}
{Katz}, J.~I. 2016, Modern Physics Letters A, 31, 1630013

\bibitem[{{Krivonos} {et~al.}(2010){Krivonos}, {Tsygankov}, {Revnivtsev},
  {Grebenev}, {Churazov}, \& {Sunyaev}}]{IBIS2010}
{Krivonos}, R., {Tsygankov}, S., {Revnivtsev}, M., {et~al.} 2010, \aap, 523,
  A61

\bibitem[{{Kumar} \& {Zhang}(2015)}]{Kumar2015}
{Kumar}, P., \& {Zhang}, B. 2015, \physrep, 561, 1

\bibitem[{{Levine} {et~al.}(1984){Levine}, {Lang}, {Lewin}, {Primini},
  {Dobson}, {Doty}, {Hoffman}, {Howe}, {Scheepmaker}, {Wheaton}, {Matteson},
  {Baity}, {Gruber}, {Knight}, {Nolan}, {Pelling}, {Rothschild}, \&
  {Peterson}}]{Levine1984}
{Levine}, A.~M., {Lang}, F.~L., {Lewin}, W.~H.~G., {et~al.} 1984, \apjs, 54,
  581

\bibitem[{{Rao} {et~al.}(2016){Rao}, {Chand}, {Hingar}, {Iyyani}, {Khanna},
  {Kutty}, {Malkar}, {Paul}, {Bhalerao}, {Bhattacharya}, {Dewangan}, {Pawar},
  {Vibhute}, {Chattopadhyay}, {Mithun}, {Vadawale}, {Vagshette}, {Basak},
  {Pradeep}, {Samuel}, {Sreekumar}, {Vinod}, {Navalgund}, {Pandiyan}, {Sarma},
  {Seetha}, \& {Subbarao}}]{Rao2016}
{Rao}, A.~R., {Chand}, V., {Hingar}, M.~K., {et~al.} 2016, \apj, 833, 86

\bibitem[{{Singh} {et~al.}(2014){Singh}, {Tandon}, {Agrawal}, {Antia},
  {Manchanda}, {Yadav}, {Seetha}, {Ramadevi}, {Rao}, {Bhattacharya}, {Paul},
  {Sreekumar}, {Bhattacharyya}, {Stewart}, {Hutchings}, {Annapurni}, {Ghosh},
  {Murthy}, {Pati}, {Rao}, {Stalin}, {Girish}, {Sankarasubramanian},
  {Vadawale}, {Bhalerao}, {Dewangan}, {Dedhia}, {Hingar}, {Katoch}, {Kothare},
  {Mirza}, {Mukerjee}, {Shah}, {Shah}, {Mohan}, {Sangal}, {Nagabhusana},
  {Sriram}, {Malkar}, {Sreekumar}, {Abbey}, {Hansford}, {Beardmore}, {Sharma},
  {Murthy}, {Kulkarni}, {Meena}, {Babu}, \& {Postma}}]{2014SPIE.ASTROSAT}
{Singh}, K.~P., {Tandon}, S.~N., {Agrawal}, P.~C., {et~al.} 2014, in
  Proceedings of SPIE, Vol. 9144, Space Telescopes and Instrumentation 2014:
  Ultraviolet to Gamma Ray, 91441S

\bibitem[{{Toma} {et~al.}(2009){Toma}, {Sakamoto}, {Zhang}, {Hill},
  {McConnell}, {Bloser}, {Yamazaki}, {Ioka}, \& {Nakamura}}]{Toma2009}
{Toma}, K., {Sakamoto}, T., {Zhang}, B., {et~al.} 2009, \apj, 698, 1042

\end{thebibliography}

\end{document}